# Growth by SR method and characterization of bis(thiourea)zinc(II) chloride single crystals

S. K. Kushwaha, N. Vijayan and G. Bhagavannarayana

*Materials Characterization Division, National Physical Laboratory, New Delhi –110 012, India*

Large size single crystals of bis(thiourea)zinc(II) chloride (BTZC), a potential nonlinear optical material, have been grown successfully by the Sankaranarayanan–Ramasamy (SR) method. Powder X-ray diffraction and Fourier transform infrared analyses confirmed the material of the grown crystal. Thermal stability was assessed by the thermogravimetric/differential thermal analysis. The high-resolution X-ray diffraction and dielectric measurements indicate that the crystal grown by the SR method has good crystalline perfection and low density of defects.

## 1. Introduction

In Thiourea, a centrosymmetric molecule with large dipole moment and ability to form an extensive network of hydrogen bonds, produces excellent noncentrosymmetric complexes with some metals such as Zn, Cd, Hg, Mn, etc, and forms organometallic crystals such as zinc thiourea sulphate (ZTS), bis(thiourea)zinc(II) chloride (BTZC), cadmium thiourea acetate (CTA) and potassium thiourea iodide (KTI) [1–4]. Most of these crystals have been grown by the slow evaporation solution growth technique (SEST) and gel technique [5–6]. Due to the inherent limitations of these techniques, the size of the crystals grown by these methods is small. Large size crystals of the above are necessary for device purposes. The recently invented Sankaranarayanan–Ramasamy (SR) method [7] finds an important place in this context. This method has already been successfully applied to grow large size crystals (with good crystalline perfection) of several important substances. In the present investigation, large size single crystals of BTZC have been grown successfully by the SR method, for the first time. The grown crystal has been characterized by carrying out powder X-ray diffraction (PXRD), Fourier transform infrared (FT-IR) spectroscopic, thermogravimetric (TG/DTA), high-resolution X-ray diffraction (HRXRD) and dielectric measurements. The results obtained are reported herein.

## 2. Experimental

The title compound was successfully synthesized by adopting the procedure reported by Rajasekaran et al. [8]. The commercially available zinc chloride and thiourea were purified by repeated recrystallization processes and the recrystallized material was used for synthesis. The reaction mechanism is shown below:

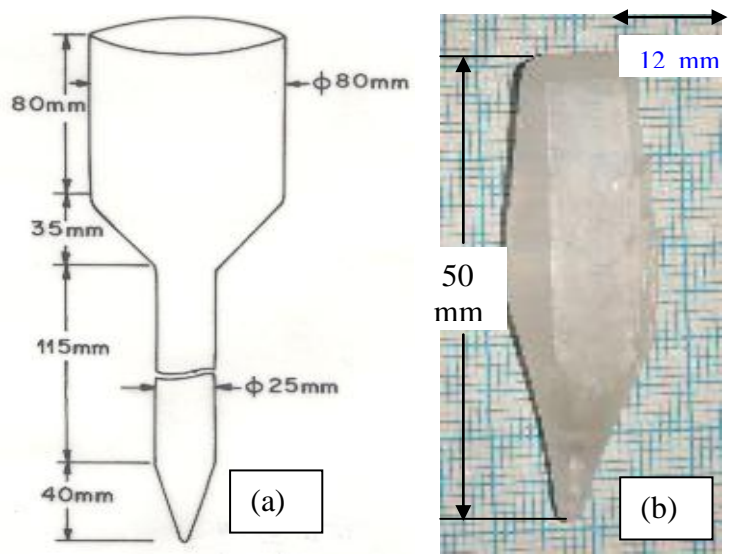

Fig. 1: (a) Schematic of SR ampoule and (b) Photograph of ZTC crystal grown by SR method.

$$ZnCl_2 + 2[CS(NH_2)_2] \rightarrow Zn[CS(NH_2)]_2Cl_2$$

The super saturated solution was prepared using the synthesized salt and after a period of 10 days a well faceted crystal having length along (which the crystal grows faster) [100] direction was harvested and it is used as a seed crystal for SR growth. The ampoule or container used in the SR methodwas made up of ordinary hollow glass tube with tapered V-shaped bottom (where the seed crystal is mounted along [100]) and U-shaped top portion to fill good amount of super saturated solution to grow bulk size crystal. The schematic of the ampoule is given in Fig. 1(a). The super saturated solution of BTZC was carefully decanted into the ampoule which was kept in CTB with a setting temperature of 25 °C. The growth conditions were closely monitored and found that the seed crystal starts increasing its size after 10 days. After a time span of 45 days a good quality single crystal of BTZC has been harvested successfully first time with size ~40 mm length and ~12mmdia. The harvested single crystal separated from the ampoule is shown in Fig.1(b) with natural facets. It may be mentioned here that by pouring some more saturated solution carefully from time to time, one can grow still bigger size crystals.

The lattice dimensions have been determined from the powder X-ray diffraction (PXRD) analysis at a scan speed of 1.2°/min. A multicrystal X-ray diffractometer (MCD) developed at National Physical Laboratory (NPL), New Delhi [9] was used to evaluate the crystalline perfection of the title compound. Before recording the diffraction curve (DC), the surface of the crystal was lapped and chemically etched in a non-preferential etchant of water and acetone mixture in 1:2 volume ratios. Such carefully prepared specimen crystals grown by both the methods were characterized. Thermal stability of SR-grown BTZC has been assessed by TG-TDA using Mettler Toledo Stare System analyzer. The cut and polished specimen of BTZC was first annealed at 100 °C to remove any possible moisture in the specimen and subjected to dielectric measurements using PSM1735 impedance analyzer.

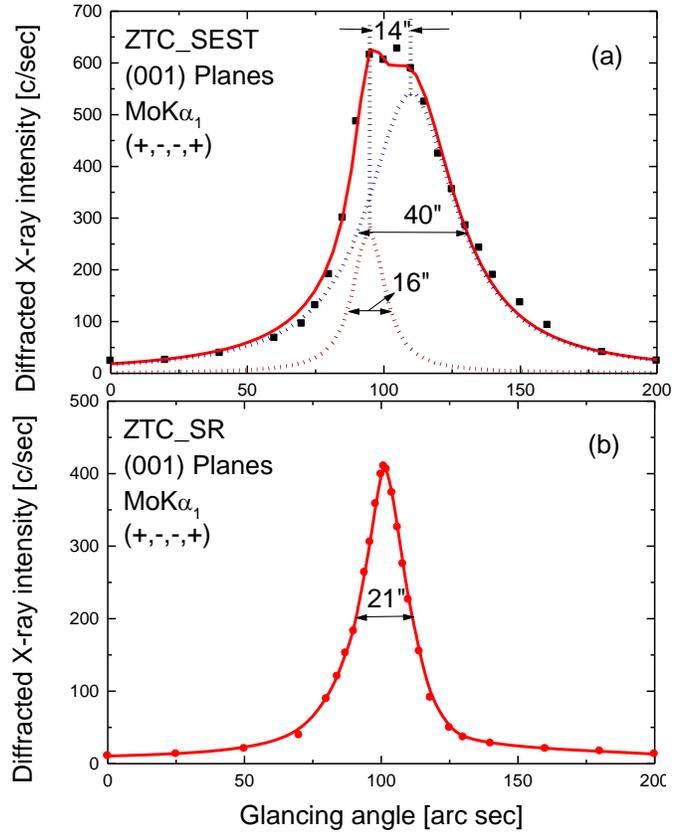

Fig. 2: High-resolution X-ray diffraction curves recorded for ZTC single crystal grown by (a) SEST and (b) SR method.

## 3. Results and discussion

From the PXRD study, it is confirmed that the grown specimen belongs to orthorhombic system having noncentrosymmetric nature with space group of Pnma. The determined cell dimensions a, b and c are respectively 13.0488, 12.5426 and 5.8925 Å. The volume of the unit cell V=964.40 Å$^3$. The observed values are consistent with the reported literature values [8]. To confirm the functional groups of the grown crystal, the FT-IR spectrum was recorded by using FT-IR Spectrometer (Nicolet-5700) in the wave number range 400–4000 cm$^{-1}$. From this analysis we found that the observed characteristic peaks belong to the synthesized compound of BTZC.

Fig. 2(a) shows the DC recorded in symmetrical Bragg geometry for a typical SEST-grown specimen for (001) diffracting planes using MoKα$_1$ radiation. On deconvolution using Lorentzian fit, it is clear that the DC contains two peaks. The additional peak is due to internal structural grain boundary [10]. The peak at 14″ away from the main peak indicates a very low angle (tilt angle ≤1′) grain boundary with a tilt angle (misorientation angle between the two crystalline regions on both sides of the structural grain boundary) of 14″. The FWHM (full width at half maximum) of the main peak and the very low angle boundary are respectively 40 and 16″. The low angular spread (∼200″) and the low values of FWHM of the DC indicate that the crystalline perfection is reasonably good.

Fig. 2(b) shows the DC for a typical SR-grown specimen recorded under identical conditions as that of Fig. 2(a). As seen in the figure, the DC is having a single and quite sharp peak. The absence of additional peak(s) in contrast to SEST-grown crystals indicates that this specimen does not contain any internal structural grain boundaries. The FWHM is 21″, which is close to that theoretically expected value of a perfect crystal according to the plane wave dynamical theory of X-ray diffraction [11]. These results show that the crystalline perfection of BTZC single crystals grown by SR method is better than that of the SEST-method without having any grain boundaries as observed in our earlier studies on benzimidazole single crystals [12].

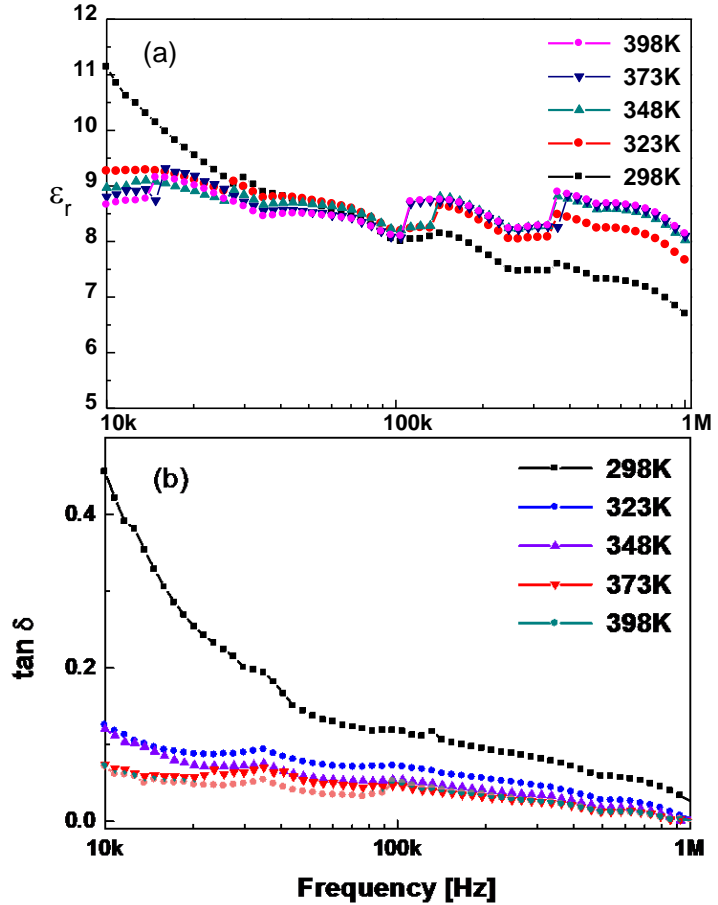

Fig. 3: Dielectric measurements at different frequencies and temperatures for ZTC crystal: (a) relative dielectric constant $\varepsilon_r$ and (b) tan$\delta$.

From the TG analysis, no weight loss was found up to 200 °C. Above 200 °C, the weight of the compound losses sharply and two endothermic peaks are observed at around 240 °C and 330 °C with a broad exothermic behavior between them as observed earlier [13]. From this analysis it may be concluded that the BTZC crystals have good thermal stability up to 200 °C and hence they can be used for device applications below this temperature. The earlier studies of SHG by Rajasekaran et al [8] confirmed that the title compound is showing second harmonic signal while passing the high intensity Nd: YAG laser.

Dielectric properties are correlated with electro-optic property of the crystals [14] particularly when they are nonconducting materials. The plots (a) and (b) in Fig. 3 shows the relative dielectric constant ($\varepsilon_r$) and loss factor (tan $\delta$) for the BTZC specimen crystals. As seen in the plot (a), the room temperature values of $\varepsilon_r$ gradually decreased as the frequency increased. However, as the temperature increases, $\varepsilon_r$ decreases up to ∼15 kHz, but increases above 100 kHz with respect to the room temperature (RT) values. These features indicate that at low frequencies, the dipolar contribution decreases where as the electronic contribution increases at high frequencies to the total polarizability of the BTZC specimen as the temperature increases. The tan $\delta$ plot [Fig. 3(b)] indicates that the loss factor decreases exponentially with frequency at RT. As the temperature increases, tan $\delta$ decreases very rapidly at low frequencies and gradually decreases at high frequencies. Low dielectric loss at higher frequencies indicate that the specimen crystal contains very low density of defects which is in tune with the HRXRD results with low FWHM value for the DC.

The good thermal stability, SHG efficiency and dielectric properties show that the title compound may be useful for variety of NLO applications.

## 4. Conclusions

Good quality, unidirectional and relatively bigger size BTZC single crystal has been grown successfully first time by SR method. Its crystalline perfection was found to be better than that of SEST-grown crystals. The crystal structure and functional groups were confirmed. The good dielectric properties observed in the present investigations along with the inherent thermal stability and relatively good SHG efficiency show that the title compound may be useful for the fabrication of devices related to NLO such as second harmonic generators, frequency doublers etc.